\documentclass[aps,groupaddress,superscriptaddress,reprint,amsmath,amssymb,graphicx
]{revtex4-1}

\usepackage{amsmath}
\usepackage{amsfonts}
\usepackage{amssymb}
\usepackage{scrextend}
\usepackage{color}
\usepackage{graphicx}
\usepackage{enumerate}
\usepackage{soul}
\usepackage{comment}
\usepackage{mathrsfs}
\usepackage{float}
\usepackage[T1]{fontenc}
\usepackage{cases}
\usepackage{ulem}
\usepackage[colorlinks,linkcolor=blue,urlcolor=blue,citecolor=blue]{hyperref}
\usepackage{appendix}

\makeatletter
\newcommand{\locallabel}[1]% #1 = label name
{\bgroup% temporary
  \edef\@currentlabel{\arabic{figure}}%
  \label{#1}%
\egroup}
\makeatother

\newcommand{\rmdt}{{\rm d}t}
\newcommand{\persta}{t_{\rm a}}
\newcommand{\traptime}{t_{\rm trap}}

\begin{document}

\title{Efficient control protocols for an active Ornstein-Uhlenbeck particle}

\author{Deepak Gupta}
\email{phydeepak.gupta@gmail.com}
\affiliation{Nordita, Royal Institute of Technology and Stockholm University, Roslagstullsbacken 23, SE-106 91 Stockholm, Sweden}
\affiliation{Department of Physics, Simon Fraser University, Burnaby, British Columbia V5A 1S6, Canada}
\affiliation{Institut f\"{u}r Theoretische Physik, Hardenbergstr. 36, Technische Universit\"{a}t Berlin, D-10623 Berlin, Germany}
\author{Sabine H.\ L.\ Klapp}
\affiliation{Institut f\"{u}r Theoretische Physik, Hardenbergstr. 36, Technische Universit\"{a}t Berlin, D-10623 Berlin, Germany}
\author{David A.\ Sivak}
\email{dsivak@sfu.ca}
\affiliation{Department of Physics, Simon Fraser University, Burnaby, British Columbia V5A 1S6, Canada}
\begin{abstract}
Designing a protocol to efficiently drive a stochastic system is an active field of research. Here we extend such control theory to an active Ornstein-Uhlenbeck particle (AOUP) in a bistable potential, driven by a harmonic trap. We find that protocols designed to minimize the excess work (up to linear-response) perform better than naive protocols with constant velocity for a wide range of protocol durations. 
\end{abstract}

\maketitle

\section{Introduction}
\label{intro}
Active matter is composed of self-propelled units that convert free energy from the environment into mechanical motion~\cite{ramaswamy2017active,Symmetry-NF}. This intrinsic self-propulsion violates detailed balance~\cite{DB_1,DB_2} and drives the system out of equilibrium~\cite{AMNEP}. Examples of such active-matter systems include flocking birds~\cite{Flock}, fish schools~\cite{Hubbard}, light-activated colloids~\cite{light-ac}, synthetic microswimmers~\cite{Dreyfus2005}, motile cells~\cite{Comelles2014}, bacteria~\cite{DB_1,berg2008coli}, and human and animal crowds~\cite{Crowd-1,Crowd-2,Crowd-3}. Researchers have uncovered fascinating behaviors in active-matter systems including jamming~\cite{Jamming}, clustering~\cite{Cluster}, and motility-induced phase separation~\cite{MIPS}. A profusion of experimental and theoretical investigations have probed their nonequilibrium nature at the single-particle level~\cite{Chen,Pietzonka_2018,Pietzonka_2019,FEPRTP,Khadka2018,Tomoyuki-Mano,toyabe,cocconi2023optimal}. 

Experiments reveal the promise of active systems for several applications~\cite{Balda2022}, such as delivering drugs to target organs~\cite{GHOSH2020100836, Gu2022}, controlling the spread of infectious microorganisms~\cite{Forgacs2022}, and developing micro-robots capable of advanced group behaviors~\cite{robo1,robo2}. Recently, researchers have focused on developing optimal schemes to transport such active particles in complex environments~\cite{Yang_2018,Liebchen_2019,Nasiri_2022}. 

Since active systems constantly dissipate energy into the environment to sustain nonequilibrium directed operations, it is of paramount importance to develop efficient driving strategies (temporal schedules for varying external control parameters) that reduce the thermodynamic costs of control~\cite{Shankar}. Examples of such control parameters include length of a polymer, stiffness and location of a particle-confining trap, and magnetic fields on spin systems~\cite{SB_review}. One way to manipulate the dynamics of nonequilibrium systems and to control the thermodynamic cost is {\it feedback}, as has been demonstrated, {\it e.g.}, for Brownian ratchet systems~\cite{FB_1,FB_2,FB_3,FB_4,FB_5,FB_6,jannik-OU}. Another promising route is to deliberately design a pre-determined protocol that does not depend on contemporary measurements of the system. Indeed, researchers have analytically obtained an optimal driving schedule (henceforth a {\it protocol}) that minimizes dissipation for a harmonically confined (passive) Brownian particle for arbitrary protocol duration~\cite{OC_PRL,OC-2}; however, far from equilibrium there is no general strategy to design a minimum-dissipation control protocol for a system diffusing in an arbitrary potential-energy landscape. 

Reference~\cite{DS-PRL} formulated a linear-response framework for such a complicated scenario to design protocols that minimize dissipation near equilibrium. This method has been used to design protocols that reduce dissipation for biomolecular systems, such as driving the F$_1$-ATPase molecular motor to synthesize ATP~\cite{optimalF1}, and driving folding and unfolding of single DNA hairpins~\cite{Tafoya5920}. Moreover, the effectiveness of this scheme has been demonstrated in numerical simulations of barrier crossing~\cite{Barrier}, rotary motors~\cite{Joseph}, Ising models~\cite{RotskoffCrooks_PRE15,RotskoffEVE_PRE17,Miranda}, and several other model systems~\cite{ZulkowskiDeWeese_PRE12,ZulkowskiDeWeese_PO13,erasure,BonancaDeffner_JCP14}.

In contrast to previous works applicable to systems in thermal equilibrium in the absence of driving~\cite{OC_PRL,OC-2,DS-PRL,optimalF1,Tafoya5920,Barrier,Joseph,RotskoffCrooks_PRE15,RotskoffEVE_PRE17,Miranda,ZulkowskiDeWeese_PRE12,ZulkowskiDeWeese_PO13,erasure,BonancaDeffner_JCP14}, here we seek efficient driving protocols that minimize the work in driving an active particle. Specifically, we drive an active Ornstein-Uhlenbeck particle (AOUP) in a double-well potential using a harmonic confinement. The AOUP is a popular active-particle model that has already been useful in investigating  motility-induced phase separation~\cite{MIPS}, glassy behavior~\cite{Glassy}, heat transport~\cite{HF}, and other active nonequilibrium behavior~\cite{STAOUP,Far-1,Far-2,Far-3}. For this system, we apply the linear-response framework~\cite{DS-PRL} to design a driving protocol.
We show that this ``designed protocol'' performs better than a naive (constant-velocity) protocol. Our analysis extends the linear-response framework (originally derived in passive close-to-equilibrium systems) to AOUPs close to a nonequilibrium stationary state. 

The rest of the paper is organized as follows. Section~\ref{setup} introduces the model. Section~\ref{theory} presents the linear-response framework. Section~\ref{results} discusses the designed protocol and its effectiveness in driving the particle over the potential-energy barrier. Section~\ref{discussions} summarizes the main results. Appendix~\ref{sec:fric_comp} compares the generalized friction obtained using the full-model defined in Eqs.~\eqref{dyn-3} and \eqref{dyn-2} (in two extreme limits of the active particle's persistence time) with that obtained using the effective model~\eqref{eff_eqn}. Appendix~\ref{sec:relax} derives the Kramers time for a passive Brownian particle. Appendix~\ref{method} discusses numerical simulation methods.

\section{Setup}
\label{setup}
\begin{figure}
    \centering
    \includegraphics[width = \columnwidth]{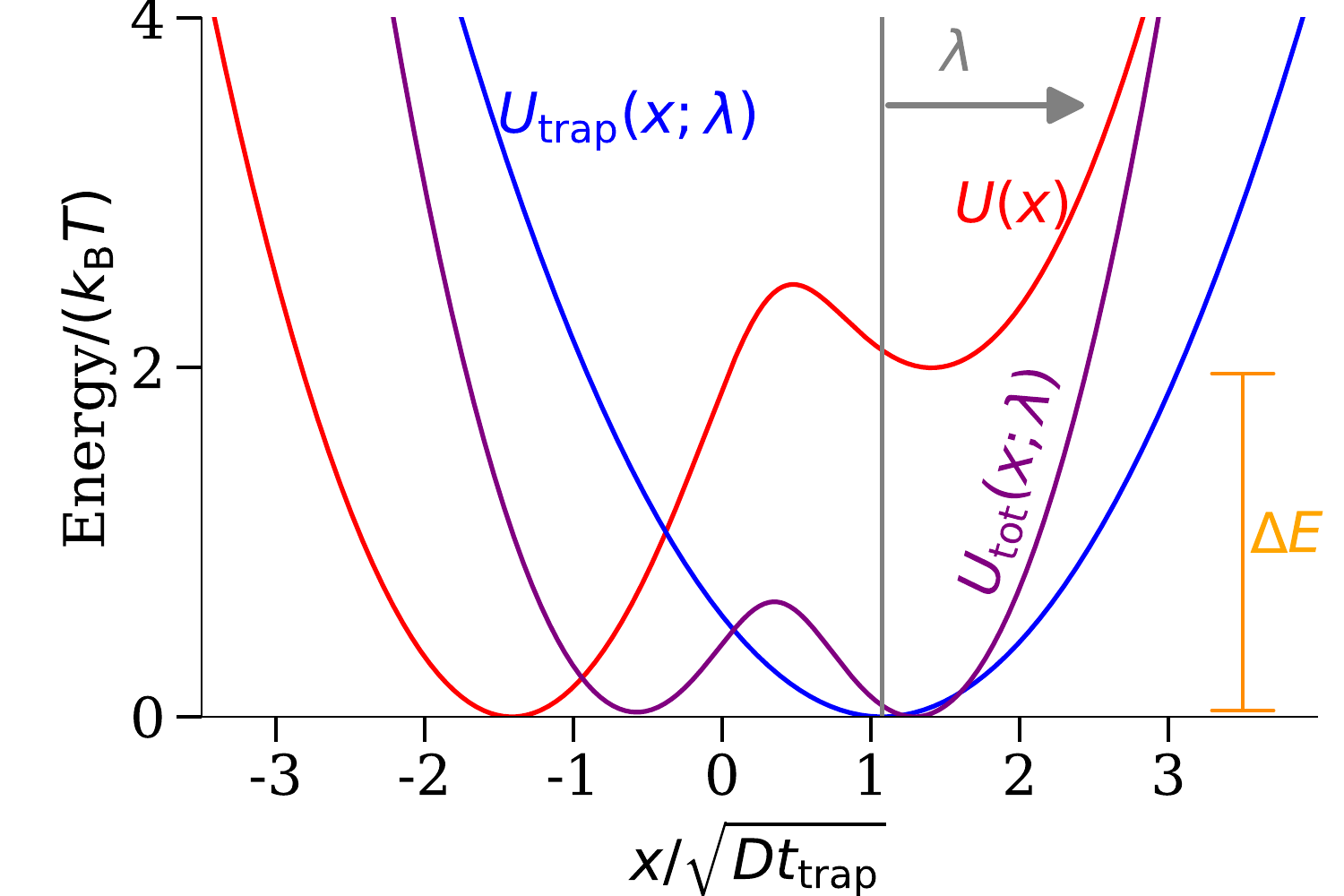}
    \caption{Model schematic. Potential-energy landscape $U(x)$, trap potential $U_{\rm trap}(x;\lambda)$, and total potential energy $U_{\rm tot}(x;\lambda)\equiv U(x) +U_{\rm trap}(x;\lambda)$, as functions of particle position $x$. Trap minimum is $\lambda=1.075\sqrt{D t_{\rm trap}}$, for trap relaxation time $t_{\rm trap}\equiv k_{\rm B}T/(D E^\ddag)$. Energy offset $\Delta E = 2~k_{\rm B}T$ between potential minima at $x=\pm x_{\rm m} = \pm 1.414\sqrt{D t_{\rm trap}}$. Here and in the following figures, the spring constant is $k=2~k_{\rm B} T/(Dt_{\rm trap})$, and the trap stiffness is $E^\ddag = k/2$.}
    \label{fig:pot}
\end{figure}

We consider an active Ornstein-Uhlenbeck particle (AOUP) coupled to a heat reservoir at temperature $T$ and confined in a one-dimensional (1D) double-well potential~\cite{Barrier} (see Fig.~\ref{fig:pot}):
\begin{align}
U(x) \equiv -\beta^{-1} \ln \bigg[e^{-\frac{\beta k}{2}(x+x_{\rm m})^2} + e^{-\beta \Delta E-\frac{\beta k}{2}(x-x_{\rm m})^2}\bigg]\ , \label{land}
\end{align}
for particle position $x$, inverse temperature $\beta \equiv (k_{\rm B}T)^{-1}$, Boltzmann's constant $k_{\rm B}$, and spring constant $k$. The double-well minima are located at $x=\pm x_{\rm m}$, and $\Delta E$ is the energy difference between these minima (see Fig.~\ref{fig:pot}). This double-well potential models a bistable system ({\it e.g.}, a DNA hairpin with folded and unfolded conformations) switching between its two metastable states (each modeled as a harmonic potential) on a time scale much faster than all other relevant system time scales~\cite{Sasa}.

In this paper, we seek a driving protocol that minimizes the work required to transport an AOUP [between the two wells of the double-well $U(x)$] using a harmonic trap
\begin{align}
U_{\rm trap}(x;\lambda) \equiv \dfrac{1}{2} E^\ddag[x-\lambda(t)]^2 \label{trap-eqn}
\end{align}
with fixed stiffness $E^\ddag$ and time-dependent minimum $\lambda(t)$. To simplify notation, we henceforth suppress its explicit time dependence. 

In the presence of the trap, the particle position $x$ evolves according to the Langevin equation
\begin{align}
\dot x = -\beta D U'_{\rm tot}(x;\lambda) + \sqrt{2D}~\eta(t) + y(t) \ ,\label{dyn-3}
\end{align}
where the dot and the prime respectively indicate a time- and a space-derivative, and $D$ the diffusion coefficient. The total potential energy $U_{\rm tot}(x;\lambda)\equiv U(x) +U_{\rm trap}(x;\lambda)$ experienced by the particle is the sum of the underlying landscape $U(x)$ and the trapping potential $U_{\rm trap}(x;\lambda)$ for a given trap minimum $\lambda$. Figure~\ref{fig:pot} shows schematics of $U(x)$, $U_{\rm trap}(x;\lambda)$, and $U_{\rm tot}(x;\lambda)$. In Eq.~\eqref{dyn-3}, the Ornstein-Uhlenbeck (OU) contribution $y(t)$ (hereafter the {\it active velocity}) to the velocity represents the fluctuating active self-propulsion and evolves according to~\cite{exp_1,exp_2}
\begin{equation}
\dot y = -\dfrac{y}{t_{\rm a}} + \dfrac{1}{t_{\rm a}}\sqrt{2D_{\rm a}}~\eta_{\rm a}(t) \ .\label{dyn-2}
\end{equation}
for the persistence time $t_{\rm a}$. We define the P\'eclet number ${\rm Pe} = D_{\rm a}/D$ as a dimensionless parameter characterizing the strength of the active noise relative to the thermal noise. In Eqs.~\eqref{dyn-3} and \eqref{dyn-2}, $\eta(t)$ is thermal noise and $\eta_{\rm a}(t)$ is ``active'' noise, each Gaussian with zero mean, i.e., $\langle \eta(t)\rangle = \langle \eta_{\rm a}(t)\rangle = 0$, and delta correlated in time, 
\begin{align}
        \langle \eta(t) \eta(t')\rangle=\langle \eta_{\rm a}(t) \eta_{\rm a}(t')\rangle = \delta(t-t')\ .
\end{align}
We further assume that the two noises are independent:
\begin{align}
\langle \eta(t) \eta_{\rm a}(t')\rangle = 0\ .
\end{align}
Angle brackets $\langle \cdots \rangle$ denote an average over both noises.

Integrating Eq.~\eqref{dyn-2} up to the long-time limit and averaging over the active noise $\eta_{\rm a}(t)$ gives $y$'s stationary-state average, $\langle y(t)\rangle=0$. In this stationary state, the temporal correlations of $y(t)$ decay exponentially~\cite{Peter_H}:
\begin{align}
\langle y(t)y(t') \rangle = \frac{D~{\rm Pe}}{t_{\rm a}}~e^{-|t-t'|/t_{\rm a}}\ . \label{exp-corr}
\end{align} 

Further, since $y(t)$ depends linearly on the Gaussian active noise $\eta_{\rm a}(t)$ [see Eq.~\eqref{dyn-2}], $y(t)$ is also Gaussian distributed with stationary-state distribution 
\begin{align}
p_{\rm ss}(y) = \frac{1}{\sqrt{2 \pi D~{\rm Pe}/t_{\rm a}}}~e^{-\frac{y^2}{2D~{\rm Pe}/t_{\rm a}}}\ .\label{st_dist_y}
\end{align}

For our later analysis, it is useful to consider two limiting cases. First, taking the limit $t_{\rm a}~\to~0$ in Eq.~\eqref{exp-corr}, $y(t)$ reduces to a zero-mean Gaussian white noise~\cite{Note1}, i.e.,
\begin{align}
\langle y(t)y(t') \rangle = 2D~{\rm Pe}~\delta(t-t')\ . \label{delta_corr_y}
\end{align} 
In the opposite limit ($t_{\rm a}\to \infty$ while holding $D$ and ${\rm Pe}$ fixed), the distribution of $y(t)$ becomes a delta-function at $y=0$, i.e, $p_{\rm ss}(y)= \delta(y)$ [see Eq.~\eqref{st_dist_y}]. To summarize, 
\begin{align}
\label{y-eqn-limits}
y(t) \to \begin{cases}
\sqrt{2D~{\rm Pe}}~\eta_{\rm a}(t)&\qquad t_{\rm a} \to 0\ ,\\
0&\qquad t_{\rm a} \to \infty \ . 
\end{cases}
\end{align}
Combining Eq.~\eqref{y-eqn-limits} with Eq.~\eqref{dyn-3}, in the stationary state for appropriate limits of $t_{\rm a}$, $x$ effectively describes the position of a passive Brownian particle with noise strength $D$ for $t_{\rm a}\to \infty$ and $D(1+{\rm Pe})$ for $t_{\rm a}\to 0$ [see Eq.~\eqref{eff_eqn}].

\section{Theory}
\label{theory}
For a single stochastic trajectory, the {\it excess work} $w^{\rm FM}_{\rm ex}$ is the difference between the work 
\begin{align}
w^{\rm FM} \equiv -\int_0^{t_{\rm dur}}~{\rm d}t~\dot\lambda~f\ , \label{traj_work1}
\end{align}
performed on the AOUP~\cite{jarzynski1997nonequilibrium} in a time-dependent protocol and its quasistatic value,
\begin{align}
W_{\rm qs} \equiv -\int_0^{t_{\rm dur}}~{\rm d}t~\dot \lambda~\langle f\rangle_\lambda\ . \label{qs-def_work}
\end{align}
Here $f \equiv -\partial_\lambda U_{\rm trap}(x;\lambda)=E^\ddag (x-\lambda)$ is the force conjugate to the control parameter $\lambda$. In Eq.~\eqref{traj_work1}, the superscript `FM' denotes the 
{\it full model}, i.e., Eqs.~\eqref{dyn-3} and~\eqref{dyn-2}. In Eq.~\eqref{qs-def_work}, angle brackets $\langle \dots \rangle_\lambda$ indicate an average at fixed trap minimum $\lambda$. Appendices~\ref{force} and \ref{sec:qs_work} respectively detail the computation of the stationary-state average force $\langle f\rangle_\lambda$ and the quasistatic work $W_{\rm qs}$.

The main quantity of interest is the ensemble-average (over initial conditions and each noise's history) excess work
\begin{align}
W^{\rm FM}_{\rm ex} \equiv \langle w^{\rm FM}_{\rm ex} \rangle =  -\int_0^{t_{\rm dur}}~{\rm d}t~\dot \lambda~\langle \delta f\rangle\ , \label{def_work}
\end{align}
where $\delta f \equiv f - \langle f\rangle_\lambda$ is the deviation of the force from its average for fixed trap minimum $\lambda$. Even though the excess work is an ensemble-average quantity, henceforth for brevity we drop explicit mention of the average. Notice that in the absence of active velocity (i.e., $y=0$), the quasistatic work equals the free-energy difference between the initial and final control-parameter values (see Appendix~\ref{sec:qs_work} and Fig.~\ref{fig:qs_work})~\cite{DS-PRL,optimalF1}.

In the absence of the moving harmonic trap~\eqref{trap-eqn}, the active system described by Eqs.~\eqref{dyn-3} and \eqref{dyn-2} approaches a nonequilibrium stationary state. The presence of the moving harmonic trap~\eqref{trap-eqn} pushes the system further from equilibrium. In view of this complicated situation, it seems challenging to find an optimal control protocol that minimizes the excess work~\eqref{def_work} for an AOUP. But a linear-response framework~\cite{DS-PRL}---originally derived for passive systems close to equilibrium---provides a framework for designing protocols that systematically reduce dissipation in a variety of systems~\cite{OC_PRL,OC-2,DS-PRL,optimalF1,Tafoya5920,Barrier,Joseph,RotskoffCrooks_PRE15,RotskoffEVE_PRE17,Miranda,ZulkowskiDeWeese_PRE12,ZulkowskiDeWeese_PO13,erasure,BonancaDeffner_JCP14}). Here, we test the applicability of this linear-response framework~\cite{DS-PRL} for a driven AOUP.

In the following, we briefly summarize the linear-response framework developed in~\cite{DS-PRL}. For a passive system (i.e., no active velocity, $y=0$) that remains close to its stationary (in this case equilibrium) state during time-dependent variation of the control parameter $\lambda$, within the linear-response approximation the instantaneous excess power (exceeding the corresponding quasistatic power) is
\begin{align}
    P^{\rm LR}_{\rm ex}(t) \approx \zeta(\lambda) \bigg(\dfrac{{\rm d}\lambda}{{\rm d}t}\bigg)^2\ .\label{ex-pow}
\end{align} 
(Here, the superscript `LR' denotes the {\it linear-response} approximation.) The time integral of this quantity over the protocol duration $t_{\rm dur}$ gives the excess work, 
\begin{align}
W^{\rm LR}_{\rm ex}= \int_0^{t_{\rm dur}}~{\rm d}t~P^{\rm LR}_{\rm ex}(t)\ . \label{lr-ex-work}
\end{align} 
In Eq.~\eqref{ex-pow}, the generalized friction coefficient $\zeta(\lambda)$ is the time-integral of the stationary-state force autocovariance:
\begin{align}
    \zeta(\lambda) \equiv \beta\int_0^\infty~\rmdt~\langle \delta f(0)~\delta f(t) \rangle_{\lambda}\ . \label{fric}
\end{align}
Appendix~\ref{sec:FAC} details computation of the force-autocovariance function at fixed trap minimum $\lambda$.

Multiplying and dividing the right-hand side~\eqref{fric} by the stationary-state force variance $\langle (\delta f)^2\rangle_{\lambda}$, we rewrite the generalized friction coefficient,
\begin{align}
   \zeta(\lambda) = \beta \langle (\delta f)^2 \rangle_{\lambda}~\tau_{\rm relax}(\lambda)\ ,
\end{align}
as the product of the force variance 
$\langle (\delta f)^2 \rangle_{\lambda}$
and the force relaxation time
\begin{align}
\tau_{\rm relax}(\lambda) \equiv \int_0^\infty~\rmdt~\dfrac{\langle \delta f(0)~\delta f(t) \rangle_{\lambda} }{\langle (\delta f)^2 \rangle_{\lambda}}\ . \label{tau-relax}
\end{align}

Following Ref.~\cite{DS-PRL}, the rate of change (hereafter {\it velocity}) of the designed protocol $\lambda^{\rm des}(t)$ that (near equilibrium) minimizes the excess work is inversely proportional to the square root of the friction coefficient, 
\begin{align}
   \dfrac{{\rm d}\lambda^{\rm des}}{{\rm d}t} = \dfrac{A^{\rm des}}{\sqrt{\zeta(\lambda)}}\ ,\label{des-prot} 
\end{align}
which differs from the constant-velocity (hereafter {\it naive}) protocol, 
\begin{align}
\dfrac{\rm d \lambda^{naive}}{\rm d t} = A^{\rm naive}\ . \label{nv-prot}
\end{align}
In Eqs.~\eqref{des-prot} and \eqref{nv-prot}, the protocol's boundary conditions $\lambda(0)=\lambda_{\rm i}$ and $\lambda(t_{\rm dur}) = \lambda_{\rm f}$ fix the constants $A^{\rm des}$ and $A^{\rm naive}$. Substituting \eqref{des-prot} in \eqref{ex-pow} yields (within the linear-response framework) a constant excess power, whereas for the naive protocol~\eqref{nv-prot}, the excess power~\eqref{ex-pow} is proportional to $\zeta(\lambda)$.

\section{Results}
\label{results}
We start by considering the energetic landscape determining the particle dynamics and the quantities entering the linear-response framework.

\begin{figure}
    \centering
    \includegraphics[width=\columnwidth, height = 20 cm]{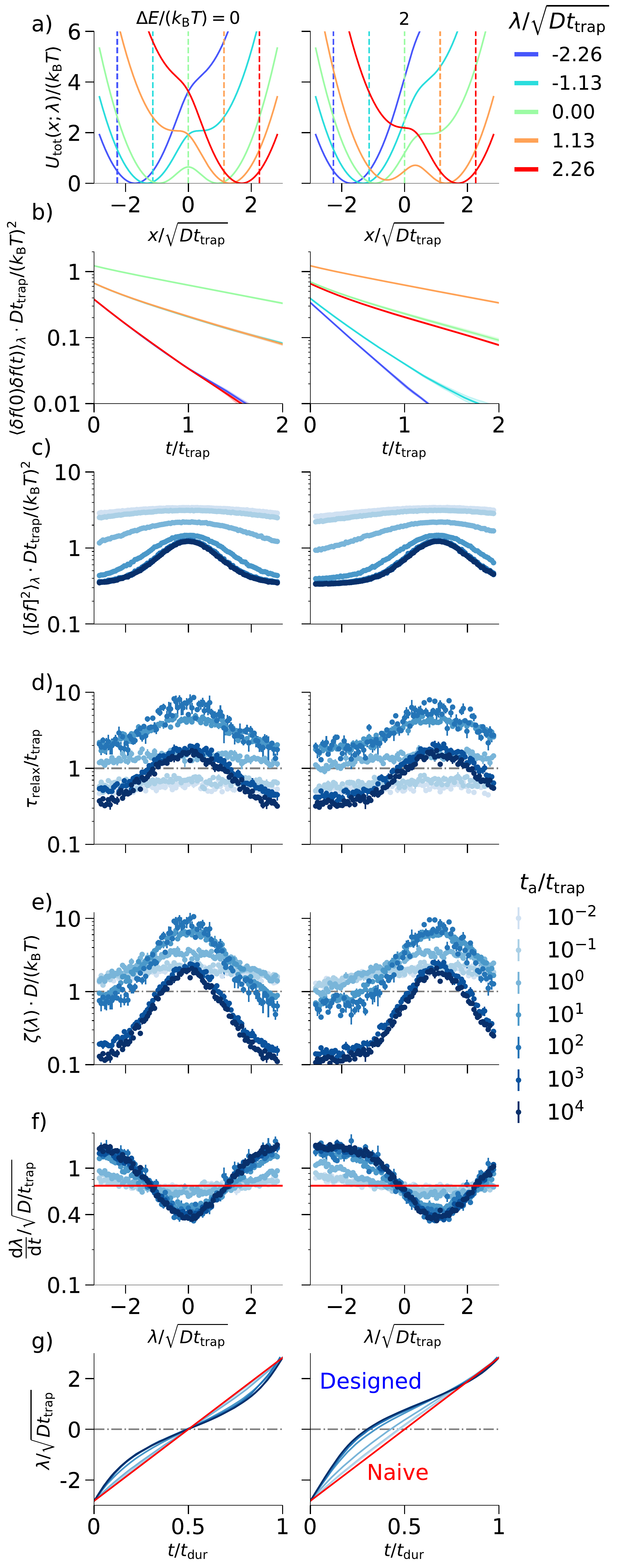}
    \caption{(a) Total potential energy as a function of particle position $x$. Vertical dashed lines: trap minimum $\lambda$. (b) Force autocovariance as a function of time, for persistence time $t_{\rm a}/t_{\rm trap}=10^4$. (c) Force variance, (d) force relaxation time, (e) generalized friction coefficient, and (f) protocol velocity, each as a function of trap minimum $\lambda$. (g) Protocol as a function of time. (f,g) Red lines: naive (constant-velocity) protocol; points/curves: designed protocols. (c-g) Blue color intensity increases with persistence time $\persta$. Here and in the following, the P\'eclet number ${\rm Pe} =5$, and vertical error bars indicate one standard error of the mean (see Appendix~\ref{method}).} 
    \label{fig:fric}
\end{figure}

Figure~\ref{fig:fric}a shows the total potential energy $U_{\rm tot}(x;\lambda)$ as a function of particle position $x$, for different trap minima $\lambda$. For each examined $\Delta E$, there is a range of $\lambda$ for which the total potential energy $U_{\rm tot}(x;\lambda)$ has two metastable states ({\it e.g.}, see $\lambda=0$ for $\Delta E = 0$).

Figure~\ref{fig:fric}b shows the force autocovariance function [determining the generalized friction coefficient~\eqref{fric}] as a function of observation time $t/t_{\rm trap}$, for different trap minima $\lambda$. The force autocovariance decays particularly slower when the total potential $U_{\rm tot}(x;\lambda)$ displays two metastable states ({\it e.g.}, for $\lambda/\sqrt{D\traptime}=0$ for $\Delta E=0~k_{\rm B}T$). For $\Delta E=0$, $U_{\rm tot}(x;\lambda)$ and $U_{\rm tot}(x;-\lambda)$ are related by a mirror reflection about $\lambda=0$ (see Fig.~\ref{fig:fric}a), thus producing identical (up to numerical sampling) force autocovariance functions; for $\Delta E \neq 0$ there is no such symmetry. 

Figures~\ref{fig:fric}c,d,e respectively display the force variance, the force relaxation time~\eqref{tau-relax}, and their product yielding the generalized friction~\eqref{fric}, each as a function of trap minimum $\lambda$. For $\Delta E = 2~k_{\rm B}T$, all these functions are asymmetric about $\lambda=0$ reflecting the asymmetry in the total potential energy landscape, $U_{\rm tot}(x;\lambda)$. For longer persistence time ($t_{\rm a}/t_{\rm trap}\gtrsim 1$), each of these quantities are maximized at a trap minimum $\lambda$ for which the total potential $U_{\rm tot}(x;\lambda)$ has two metastable states (see Fig.~\ref{fig:fric}a). However, for short persistence time ($t_{\rm a}/t_{\rm trap} \ll 1$), they are almost independent of the trap minimum $\lambda$. This is because in this limit the active velocity $y(t)$ behaves as Gaussian white noise~\eqref{delta_corr_y}, producing a higher effective diffusion coefficient $D(1+{\rm Pe})$ than the passive Brownian particle [see Eq.~\eqref{eff_eqn}]. Thus, at ${\rm Pe} = 5$ the effective temperature experienced by the AOUP is $6~k_{\rm B}T$ (six times larger than the passive Brownian particle), dominating the $\sim$1$~k_{\rm B}T$ height of the total potential's barrier (Fig.~\ref{fig:fric}a). Fig.~\ref{fig:comp_fric} shows agreement of the generalized friction coefficient obtained at extreme values of persistence time with that obtained using the effective dynamics~\eqref{eff_eqn}.

Figure~\ref{fig:fric}f shows the designed protocol velocity defined according to~\eqref{des-prot}, as a function of trap minimum $\lambda$. The system is driven slower where the generalized friction is higher in order to harness thermal fluctuations to overcome the total potential's barrier between the two metastable states, thereby reducing the excess work.

Integrating the protocol's velocity with respect to time gives the designed protocol, that is, the optimal trajectory of the trap minimum $\lambda$, as a function of time (Fig.~\ref{fig:fric}g). Since for $t_{\rm a}/t_{\rm trap} \ll 1$ the effect of the total potential's barrier is negligible (Figs.~\ref{fig:fric}c,d,e), the naive and designed protocols are indistinguishable.

In the following, we use the naive and designed protocols to compute the excess work and normalized flux using the full model described by dynamics~\eqref{dyn-3} and \eqref{dyn-2} as functions of protocol duration (see Appendix~\ref{sim-ex-wk} for the numerical simulation method). 

\begin{figure} 
    \centering
    \includegraphics[width = \columnwidth]{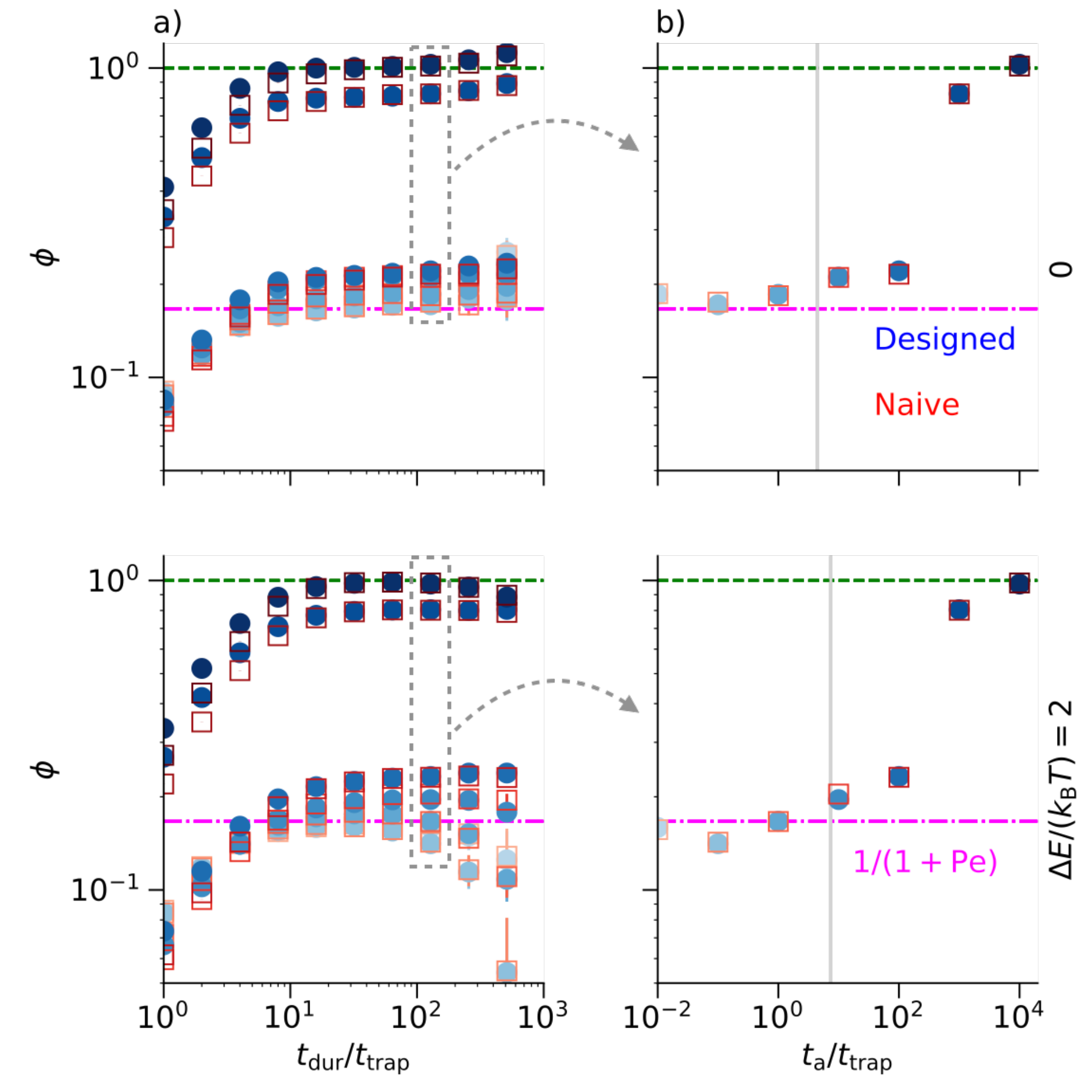}
    \caption{The linear-response accuracy $\phi\equiv W_{\rm ex}^{\rm FM}/W_{\rm ex}^{\rm LR}$, the ratio of the works under the full-model~\eqref{def_work} (FM) and the linear-response (LR) approximation~\eqref{lr-ex-work}~\cite{Note2}, a) as a function of protocol duration $t_{\rm dur}$ for various persistence times $t_{\rm a}$, and b) as a function of persistence time $t_{\rm a}$. Panel b) shows boxed data from panel a). Vertical gray lines mark the Kramers time for the passive Brownian particle, $t_{\rm K}/\traptime = 4.46\dots$ and $7.43\dots$ for $\Delta E = 0$ and $2~k_{\rm B}T$, respectively (see Appendix~\ref{sec:relax}). Red: naive; blue: designed.  Horizontal green dashed and pink dot-dashed lines respectively show $\phi=1$ and $\phi=1/(1+{\rm Pe})$. Color intensity increases with persistence time $\persta$ (see Fig.~\ref{fig:fric}).}
    \label{fig:ratio_NS_LR}
\end{figure}

\begin{figure}
    \centering
    \includegraphics[width = \columnwidth]{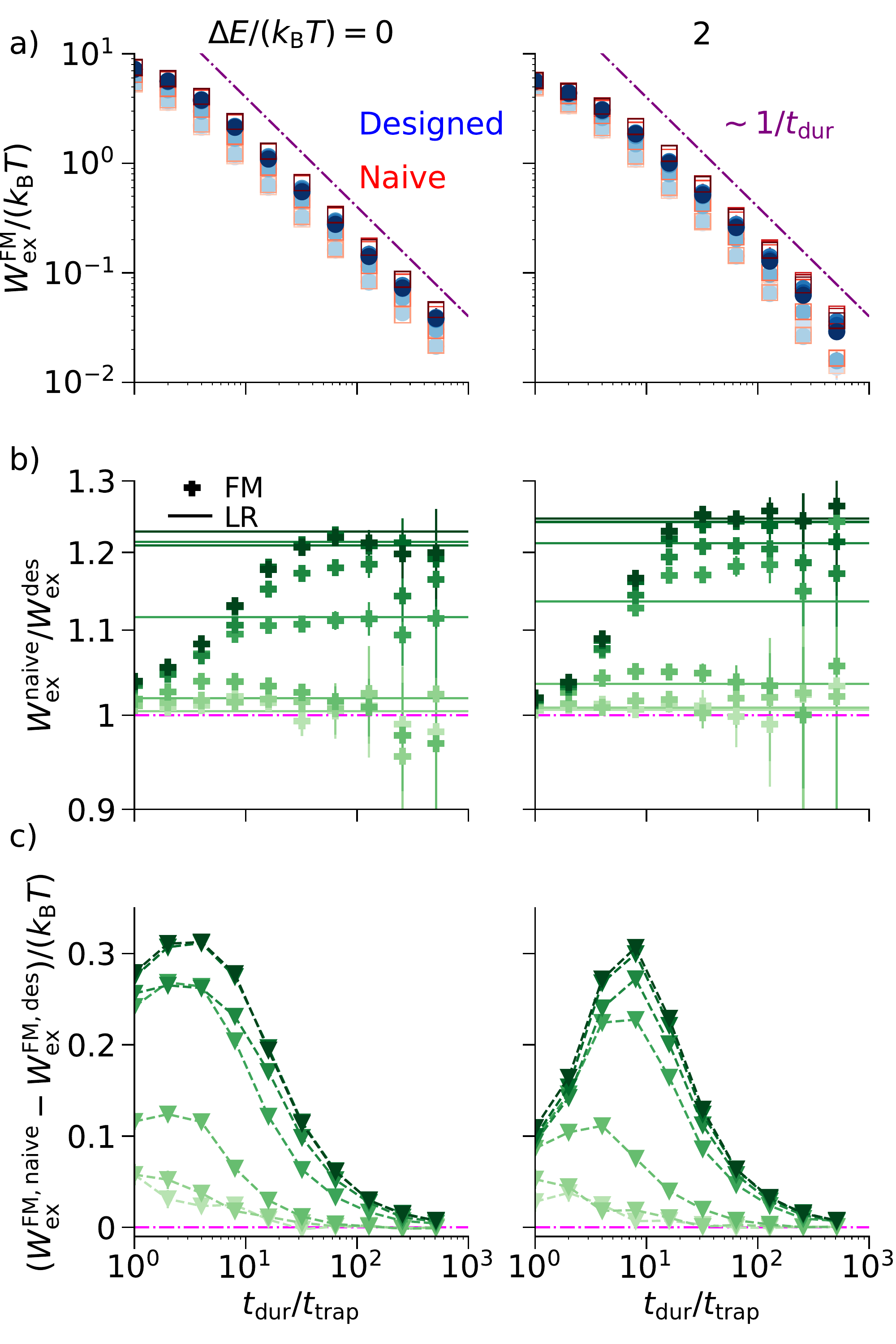}
    \caption{Excess work $W_{\rm ex}$ as a function of protocol duration $t_{\rm dur}$. a) Full-model (FM) excess work for naive (red) and designed (blue) protocols. b) Ratio of naive and designed excess works. Symbols: full-model. Horizontal lines: linear-response (LR) approximation~\cite{Note2}. c) The difference of the naive and designed full-model excess works. Dashed curves are a guide to the eye. Throughout, color intensity increases with persistence time $t_{\rm a}$ (see Fig.~\ref{fig:fric}).}
    \label{fig:excess_works}
\end{figure} 

We start by assessing the accuracy of the linear-response framework by comparing the true excess work~\eqref{def_work} using the full model~\eqref{dyn-3} and \eqref{dyn-2} and the linear-response approximation~\eqref{lr-ex-work} in the slow-driving (long-duration)
regime. Figure~\ref{fig:ratio_NS_LR}a shows the ratio $\phi \equiv W_{\rm ex}^{\rm FM}/W_{\rm ex}^{\rm LR}$, for both naive and designed protocols, as a function of protocol duration. By definition, $\phi$ quantifies the accuracy of the linear-response approximation~\cite{Note2}: $\phi = 1$ indicates complete accuracy. For each value of $t_{\rm a}$, $\phi$ approaches a constant value (up to numerical sampling) in the limit of long duration. These values appear to be independent of the protocol type and the energy shift $\Delta E$.

Figure~\ref{fig:ratio_NS_LR}b shows that this linear-response accuracy $\phi$ asymptotes to unity for $\persta\gg t_{\rm K}$, and to $1/(1+{\rm Pe})$ for $\persta\ll t_{\rm K}$, where $t_{\rm K}$ is an average Kramers time obtained for the passive Brownian case, see Appendix~\ref{sec:relax} for details. In the limit $t_{\rm a}\to \infty$, the OU contribution $y(t)$ to the velocity effectively vanishes and the system can be described by a (passive) Brownian dynamics with unchanged temperature $k_{\rm B}T$~[see Eq.~\eqref{eff_eqn} for ${\rm Pe}=0$]. In the opposite limit of $t_{\rm a}\to 0$, $y(t)$ effectively becomes an additional Gaussian white noise which combines with the Gaussian thermal white noise $\eta(t)$ to give white noise with total effective strength $D(1+{\rm Pe})$; therefore, the system can be described by a Brownian dynamics with effective temperature $k_{\rm B}T(1+{\rm Pe})$ [see Eq.~\eqref{eff_eqn}]. Figure~\ref{fig:ratio_NS_LR}b also displays the crossover of $\phi$ from $1/(1+{\rm Pe})$ to 1 as a function of persistence time $\persta$.

Figure~\ref{fig:excess_works}a shows the full-model naive and designed excess works~\eqref{def_work}, each as a function of protocol duration. At long protocol duration, the system mostly follows the trap and remains close to its stationary state during the entire protocol, so the work performed on the AOUP approaches its quasistatic value, i.e., $W^{\rm FM}_{\rm ex}\to 0$ as $t_{\rm dur}\to \infty$. We observe that this excess work (for both naive and designed protocols) decays as $\sim t^{-1}_{\rm dur}$.  

Figure~\ref{fig:excess_works}b compares the ratio of full-model naive and designed excess works~\eqref{def_work} with its linear-response approximation~\cite{Note2}, as a function of protocol duration. The linear-response approximation is more accurate at longer protocol durations. For short persistence time ($\persta/\traptime \ll 1$), the effect of the total potential's barrier on the AOUP is negligible, so the naive and designed protocols are similar (Fig.~\ref{fig:fric}g), and thus this ratio is approximately unity. Away from this limit (i.e., $\persta/\traptime \gtrsim 1$), the designed excess work is lower than the naive for a considerable range of protocol durations. We emphasize that in contrast to the absolute value of excess work (see Fig.~\ref{fig:ratio_NS_LR}), the excess-work ratio is independent of the linear-response accuracy $\phi$, signaling the applicability of the linear-response framework~\cite{DS-PRL} for the AOUP.

Figure~\ref{fig:excess_works}c shows the difference of the full-model naive and designed excess works as a function of protocol duration. For slower protocols, both naive and designed excess works decay to zero (see Fig.~\ref{fig:excess_works}a); therefore, their difference also approaches zero. For vanishing duration, all protocols produce the same excess work, so this difference again vanishes. For intermediate durations, this difference attains a maximum value indicating a protocol duration for which the designed protocol has greatest advantage over the naive protocol. The advantage of the designed protocol over the naive one is expected to be greater for greater range of variation of the generalized friction coefficient~\cite{Barrier,optimalF1}; in our model this can be achieved by a longer persistence time (Fig.~\ref{fig:fric}e) or higher energy barrier [higher $k$ in \eqref{land}].

Finally, we calculate the total flux induced by driving,
\begin{align}
\bar J \equiv \dfrac{1}{\langle x \rangle\big|_{\lambda_{\rm f}}-\langle x \rangle\big|_{\lambda_{\rm i}}}\int_0^{t_{\rm dur}}~{\rm d}t~\langle \dot x \rangle\ , \label{jbar}
\end{align}
normalized by a prefactor quantifying the distance between the mean particle positions at the control-parameter endpoints. 

\begin{figure}
    \centering 
    \includegraphics[width = \columnwidth]{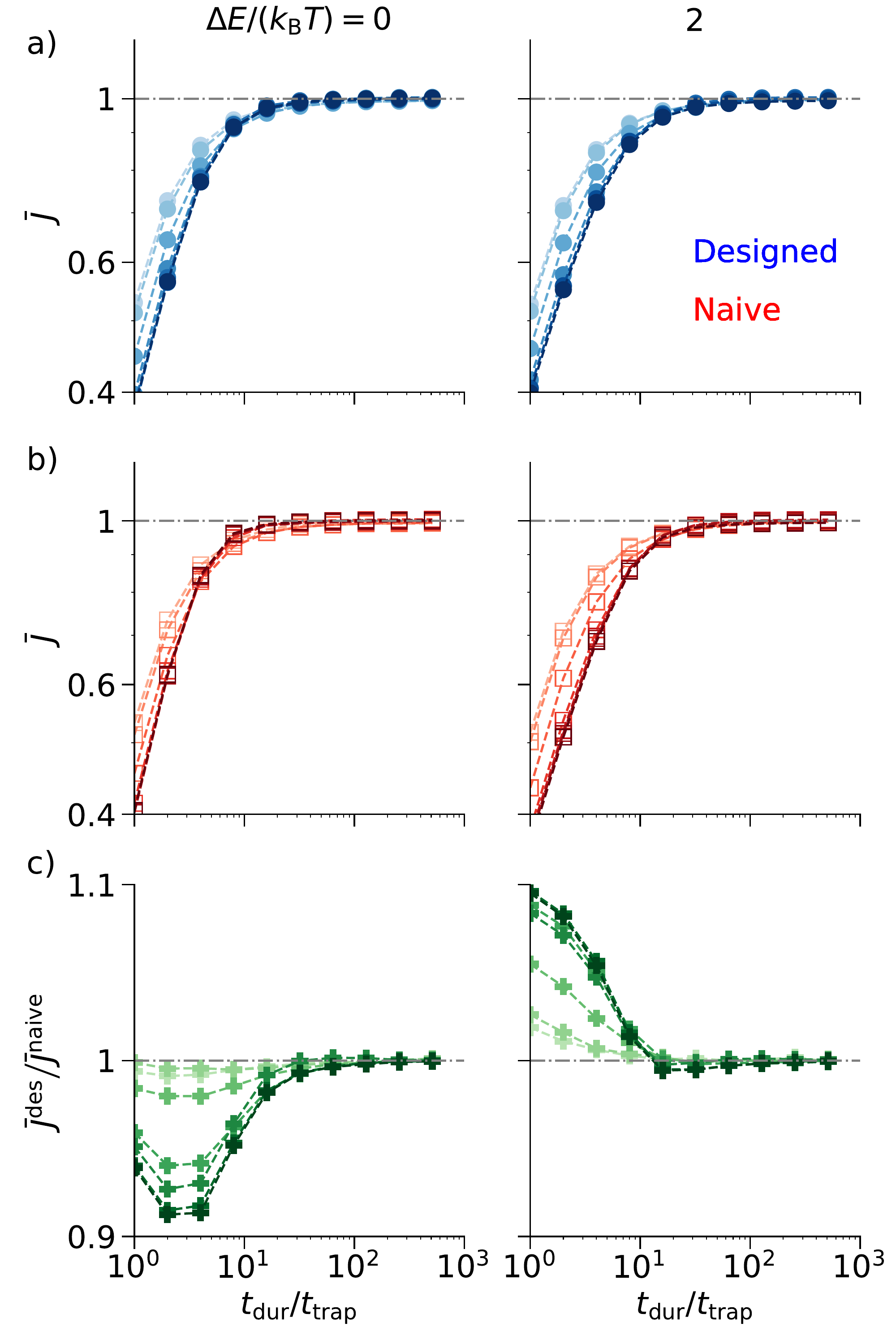}
    \caption{Normalized flux $\bar J$ as a function of protocol duration,
    for
    a) designed
    protocols, 
    b) naive
    protocols, and c) their ratio. Dashed lines are a guide to the eye. 
    Color intensity increases with persistence time $t_{\rm a}$ (see Fig.~\ref{fig:fric}).}
    \label{fig:flux}
\end{figure}

Figures~\ref{fig:flux}a,b show that 
$\bar J$ increases with protocol duration, reaching unity for longer durations (with only minor differences between protocol types), indicating successful transport over the potential-energy barrier. At shorter protocol durations,
both designed and naive fluxes increase with decreasing persistence time $t_{\rm a}$: the higher effective temperature $k_{\rm B}T(1+{\rm Pe})$ in this limit
makes it easier to cross the barrier.

Figure~\ref{fig:flux}c displays the ratio of designed flux to naive flux. 
For longer protocol durations, this ratio asymptotes to unity. For shorter durations, the designed flux is higher than naive for $\Delta E = 2~k_{\rm B}T$,
and vice versa for $\Delta E = 0$.

\section{Discussion}
\label{discussions}
In this paper, we designed a driving protocol to transport an AOUP in a 1D nonlinear potential-energy landscape using a harmonic trap. Our analysis reveals that the designed protocol obtained using the linear-response framework~\cite{DS-PRL} requires less work than the naive protocol for a considerable range of protocol durations. Moreover, at intermediate duration the work savings are maximized. Thus the linear-response result in~\cite{DS-PRL} (previously applied to systems without intrinsic activity) can be usefully extended to an AOUP.

This study opens a new research avenue investigating
the applicability of the linear-response framework to construct analogous minimum-dissipation control protocol for other active-particle systems, such as active Brownian particles~\cite{Khatami2016,Basu_u,Yasuda} and run-and-tumble particles~\cite{malakar2018steady,FEPRTP}. 
We expect that our methodology can also be extended to an active system~\cite{active2016} involving a periodic potential, such as the F$_1$-ATPase molecular motor~\cite{optimalF1}. An interesting question for future studies (but beyond the scope of this paper) is how the linear-response results, including the generalized friction coefficient $\zeta(\lambda)$, depend on  P\'eclet number ${\rm Pe}$.
Further, we emphasize that our results can be tested in an experiment driving the extension of single DNA hairpins~\cite{Tafoya5920}, but now the beads attached to the hairpin's ends experience an additional OU noise generated by electrodes coupled to a resistor and an amplifier~\cite{jannik-OU} (see Refs.~\cite{exp_1,exp_2} for other methods to generate OU noise). 

\section*{Acknowledgments}
D.G.\ acknowledges the Nordita fellowship program. Nordita is partially supported by Nordforsk. D.G.\ and S.H.L.K.\ gratefully acknowledge support from the Deutsche Forschungsgemeinschaft (DFG, German Research Foundation), Project No. 163436311-SFB 910. D.A.S.\ is supported by a Natural Sciences and Engineering Research Council of Canada (NSERC) Discovery Grant RGPIN-2020-04950 and a Tier-II Canada Research Chair CRC-2020-00098, and was enabled in part by support provided by BC DRI Group and the Digital Research Alliance of Canada~\cite{Note3}.

\renewcommand{\thefigure}{\arabic{figure}}
\renewcommand{\appendixname}{Appendix}
\appendix

\section{Generalized friction coefficient: comparison with effective dynamics}
\label{sec:fric_comp}
Figure~\ref{eff_eqn} shows the agreement of the generalized friction coefficient $\zeta(\lambda)$ for two extreme values of persistence time, $\persta/\traptime = 10^{-2}$ and $10^4$, at fixed P\'eclet number ${\rm Pe} = 5$ (see Fig.~\ref{fig:fric}e), with that obtained from the effective dynamics [substituting Eq.~\eqref{y-eqn-limits} in \eqref{dyn-3}]:
\begin{align}
\dot x = -\beta D U'_{\rm tot}(x;\lambda) + \sqrt{2D(1+{\rm Pe})}~\eta_{\rm eff}(t)\ , \label{eff_eqn} 
\end{align}
at ${\rm Pe} = 5$ and $0$ [corresponding respectively to the first and second lines of Eq.~\eqref{y-eqn-limits}]. Notice that in Eq.~\eqref{eff_eqn} $\eta_{\rm eff}(t)=\eta(t)$~\eqref{dyn-3} and ${\rm Pe}=0$ for the second line of Eq.~\eqref{y-eqn-limits}. $\eta_{\rm eff}(t)$ is Gaussian noise with zero mean, $\langle\eta_{\rm eff}(t)\rangle = 0$, and delta-correlation in time:
\begin{align}
\langle \eta_{\rm eff}(t)\eta_{\rm eff}(t')\rangle = \delta(t-t')\ .
\end{align}
As expected, the effective dynamics reproduce the generalized friction coefficient of the full dynamics in both limits.

\begin{figure} 
    \centering
    \includegraphics[width = \columnwidth]{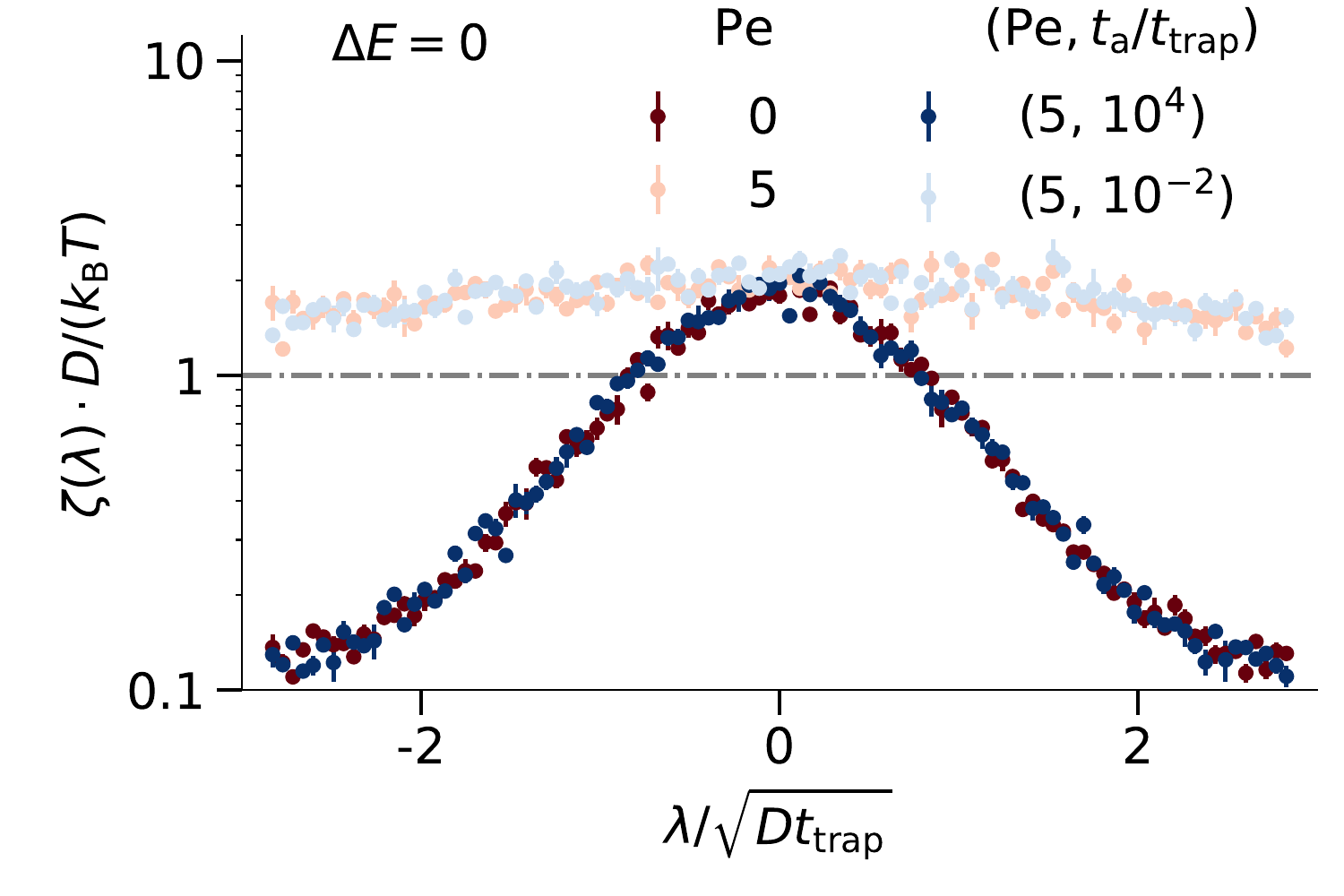}
    \caption{Generalized friction coefficient $\zeta(\lambda)$, as a function of trap minimum $\lambda$. Blue: full model~\eqref{dyn-3} from Fig.~\ref{fig:fric}e for two extreme values of persistence times, $t_{\rm a}/t_{\rm trap} = 10^{-2}$ and $10^4$. Red: effective dynamics~\eqref{eff_eqn} obtained from combining Eq.~\eqref{y-eqn-limits} and~\eqref{dyn-3}.}
    \label{fig:comp_fric}
\end{figure}

\section{Kramers time for passive Brownian particle}
\label{sec:relax}
Here we calculate the Kramers time for the passive Brownian particle, namely the characteristic time for the passive Brownian particle to transition from one well to another. This gives the vertical lines in Fig.~\ref{fig:ratio_NS_LR}b. 

The Kramers {\it rate} for the diffusion of the passive Brownian particle (dynamically evolving according to \eqref{eff_eqn} with ${\rm Pe} = 0$) to the location $\langle x \rangle_{\lambda_{\rm f}}$, starting from $\langle x \rangle_{\lambda_{\rm i}}$ is~\cite{van1992stochastic}
\begin{align}
\kappa(\lambda) \equiv \bigg[\dfrac{1}{D} \int_{\langle x \rangle _{\lambda_{\rm i}}}^{\langle x \rangle _{\lambda_{\rm f}}}~{\rm d}y~e^{\beta U_{\rm tot}(y;\lambda)}\int_{-\infty}^{y}~{\rm d}z~e^{-\beta U_{\rm tot}(z;\lambda)}\bigg]^{-1}\ ,
\end{align}
for fixed trap minimum $\lambda$. This gives the mean number of such transitions [$\langle x \rangle _{\lambda_{\rm i}} \rightarrow \langle x \rangle _{\lambda_{\rm i}}$] per unit time. We define the Kramers time for the passive Brownian particle as the inverse of the average of this Kramers rate over all fixed trap minima from $\lambda_{\rm i}$ to $\lambda_{\rm f}$:
\begin{align}
t_{\rm K} \equiv \left[\dfrac{1}{N+1}\sum_{\ell = 0}^{N}\kappa(\lambda_\ell)\right]^{-1}\ ,
\end{align}
for $\lambda_0\equiv \lambda_{\rm i}$, $ \lambda_N \equiv \lambda_{\rm f}$, $N\equiv \dfrac{\lambda_{\rm f}-\lambda_{\rm i}}{\Delta \lambda}$ and trap-minimum bin width $\Delta \lambda$. 

So when $\persta\gg t_{\rm K}$, the AOUP experiences an OU velocity that is relatively constant on the characteristic timescale for a transition (of the passive Brownian particle); conversely, when $\persta\ll t_{\rm K}$, the effect of the OU velocity on barrier crossing is effectively that of white noise.

\section{Numerical simulation methods}
\label{method}

\subsection{Force autocovariance}
\label{sec:FAC}
To compute the force autocovariance (Fig.~\ref{fig:fric}b), we discretize the Langevin equations~\eqref{dyn-3} and \eqref{dyn-2} (for each fixed $\lambda$) to first order in the discretization time $\Delta t$ and evolve the dynamics iteratively for $1\leq j\leq t/\Delta t$, where $t$ is the observation time:
\begin{subequations}
\begin{align}
x_j &= x_{j-1} - \beta D U'_{\rm tot}(x_{j-1}|\lambda)\Delta t \nonumber \\
&\quad + \sqrt{2D\Delta t}~\eta_{j-1} + y_{j-1}\Delta t \ ,\label{dyn-5} \\
y_i &= y_{j-1} - \dfrac{y_{j-1}}{\persta}\Delta t + \dfrac{1}{\persta}\sqrt{2D~{\rm Pe}~\Delta t}~\eta_{{\rm a}, j-1}\ \label{dyn-6}.
\end{align}
\end{subequations}
$\eta_{j}$ and $\eta_{{\rm a},j}$ are standard independent Gaussian random variables at the $j$th time increment, with zero mean and covariances 
\begin{align}
\langle \eta_j~\eta_k \rangle &= \langle \eta_{{\rm a},j}~\eta_{{\rm a},k}\rangle = \delta_{j,k}\\
\langle \eta_j~\eta_{{\rm a},k} \rangle &= 0\ , 
\end{align}
for Kronecker delta $\delta_{j,k}$. For a given initial condition $x_0$ and $y_0$, we generate a time-series of the force $f_j = (x_j-\lambda)E^\ddag$ at fixed trap minimum $\lambda$. To remove any dependence on initial condition, we discard the initial portion of the trajectory ($\sim$8 times the largest force relaxation time $\tau_{\rm relax}$ (see Fig.~\ref{fig:fric}d), $\sim$80 times the trap relaxation time $\traptime$), and use the remaining time-series to compute the force autocovariance $\langle \delta f_0~\delta f_j \rangle_\lambda$. We generate three independent force trajectories, each of length $t/\traptime = 1.6\cdot 10^4$, and average over the three resulting force autocovariances.

\subsection{Stationary-state average force}
\label{force}
\begin{figure}[!h] 
    \centering
    \includegraphics[width = \columnwidth]{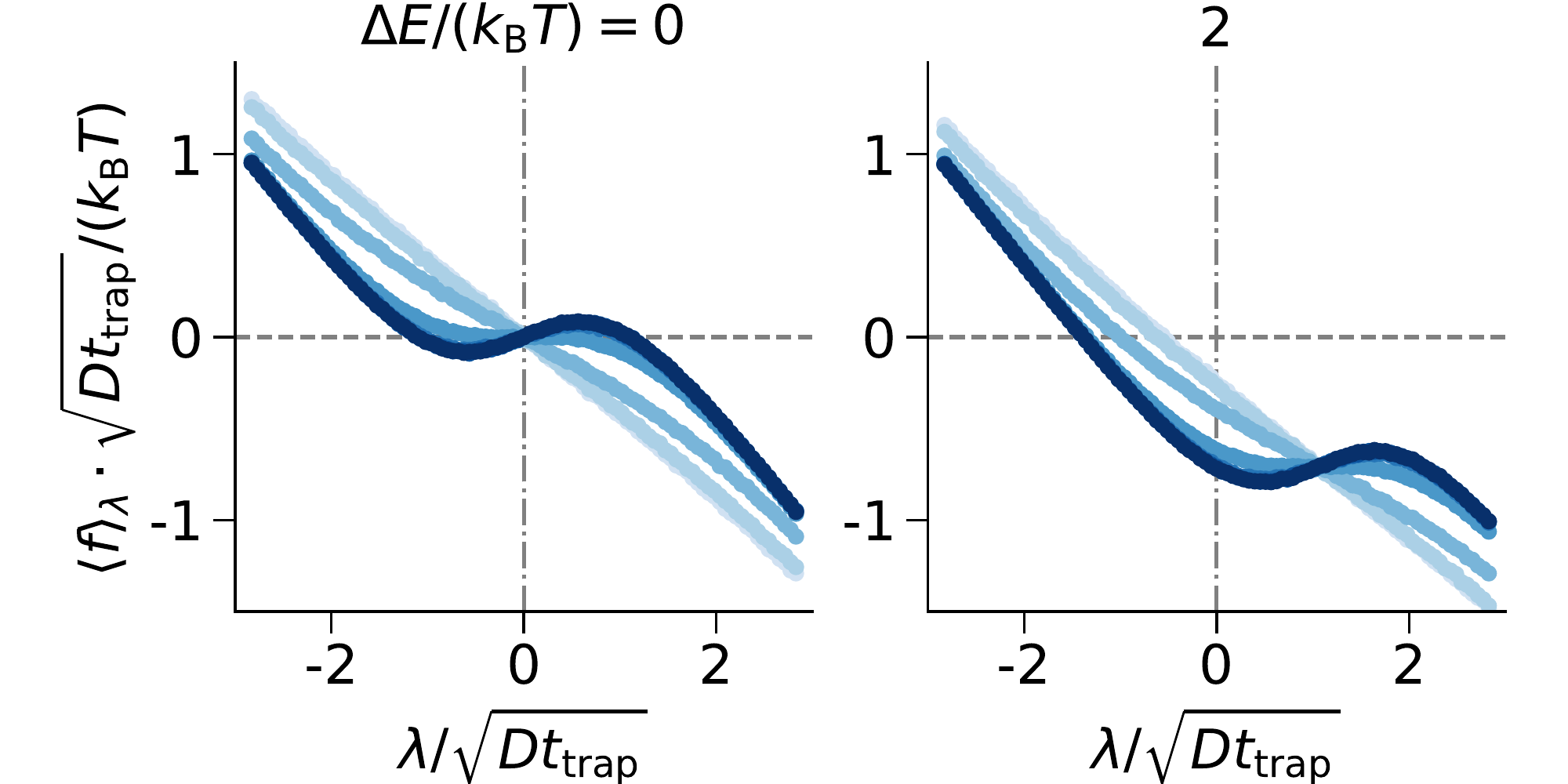}
    \caption{Stationary-state average force $\langle f\rangle_\lambda$ as a function of trap minimum $\lambda$. Color intensity increases with persistence time $\persta$ (see Fig.~\ref{fig:fric}). Error bars corresponding to one standard error of mean are smaller than the symbol.}
    \label{fig:force}
\end{figure}

We evolve the discretized Langevin equations~\eqref{dyn-5} and \eqref{dyn-6} from a fixed initial condition ($x_0 = \lambda,~y_0 = 0$) up to time $t/\traptime=8\cdot 10^2$ (ensuring the stationarity of the joint probability density function of $x$ and $y$) and compute the force experienced by the particle,
\begin{align}
f = (x-\lambda) E^\ddag \ ,
\end{align}
using the particle's position $x$ at the final time-step. Figure~\ref{fig:force} displays the stationary-state average force computed by averaging over $\mathcal{N}_{\rm R} = 10^5$ realizations for each fixed trap minimum $\lambda$. For the smallest $\persta$, the force decreases linearly as $\lambda$ increases, and appears unaffected by the barrier of the total~potential $U_{\rm tot}(x;\lambda)$.

\subsection{Quasistatic work}
\label{sec:qs_work}
We compute the discretized version of the quasistatic work [see Eq.~\eqref{qs-def_work}]: 
\begin{align}
W_{\rm qs} = -\sum_i \langle f \rangle_{\lambda_i} \Delta \lambda\ ,\label{ss-work}
\end{align}
where Fig.~\ref{fig:force} shows the average force $\langle f \rangle_{\lambda}$. 

\begin{figure}
    \centering
    \includegraphics[width = \columnwidth]{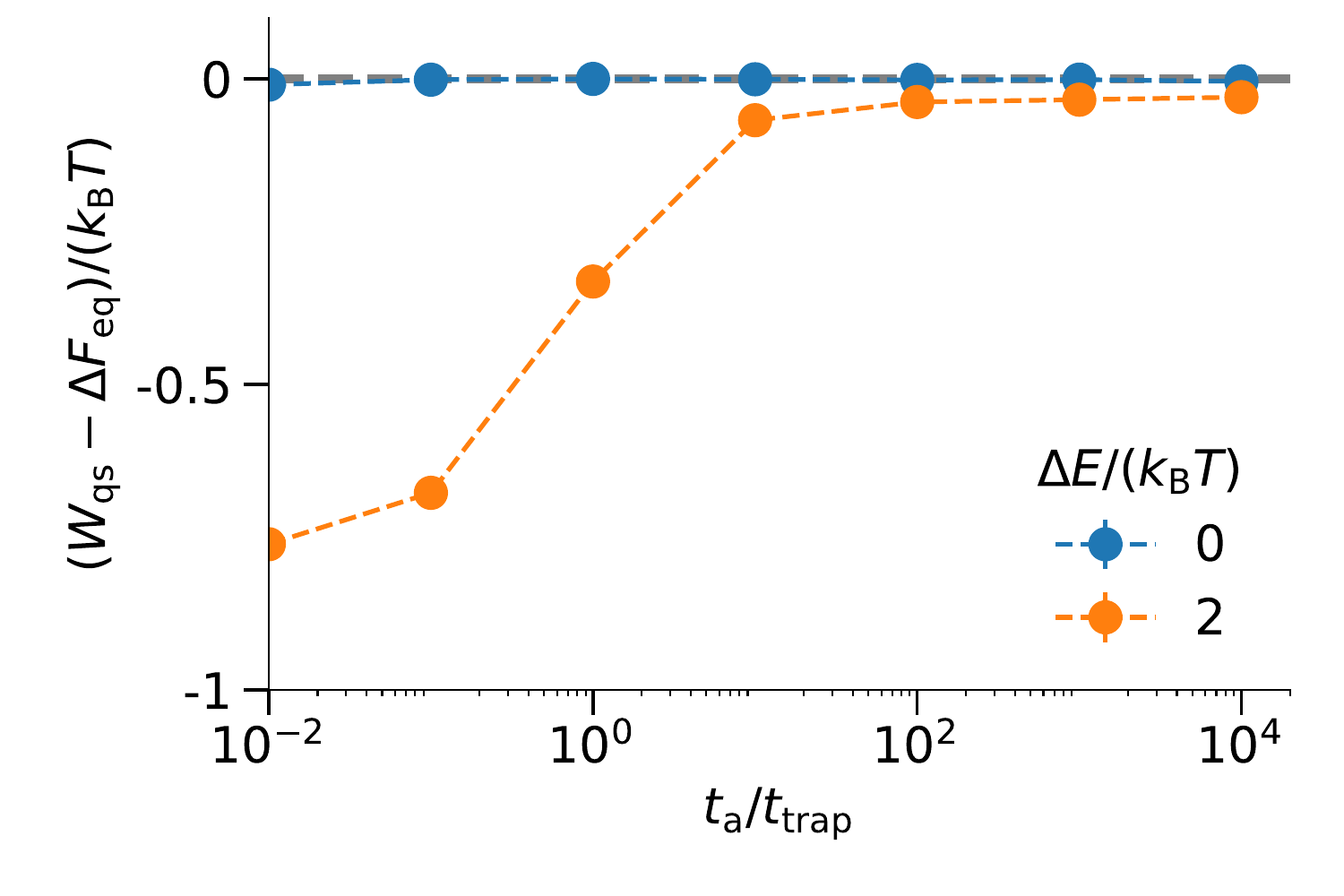}
    \caption{Difference of quasistatic work $W_{\rm qs}$ and equilibrium free-energy difference $\Delta F_{\rm eq}=\Delta E$ (for no active velocity, $y=0$), as a function of persistence time $\persta$. Dashed lines are a guide to the eye. Error bars indicating one standard error of the mean are smaller than the symbol.}
    \label{fig:qs_work}
\end{figure}

Figure~\ref{fig:qs_work} displays the difference of quasistatic work and equilibrium free-energy difference, as a function of persistence time. For longer persistence time, this difference decreases, since the system can be approximated by the effective dynamics~\eqref{eff_eqn} for ${\rm Pe}=0$, reproducing the system's passive behavior.

\subsection{Excess work}
\label{sim-ex-wk}
We use the discretized Langevin equations~\eqref{dyn-5} and \eqref{dyn-6}, interleaved with substeps that discretely update $\lambda$ according to either a naive or designed protocol (Fig.~\ref{fig:fric}g). For each trajectory, we compute the external work as the energy change due to changes of $\lambda$:
\begin{align}
w^{\rm FM} = \sum_{j = 1}^{t_{\rm prot}/\Delta t} [U_{\rm trap}(x_{j-1}|\lambda_{j})-U_{\rm trap}(x_{j-1}|\lambda_{j-1})]\ .~\label{st_work}
\end{align}
The initial condition $(x_0,y_0)$ is drawn from the (numerically computed) stationary-state distribution $\rho_{\rm ss}(x_0,y_0)$ for $\lambda_{\rm i}\equiv \lambda_0 = -2.824\sqrt{Dt_{\rm trap}}$. We simulate a range of protocol durations, with the shortest duration of $\traptime$ $\sim$5 times the smallest force relaxation time $\tau_{\rm relax}$ (Fig.~\ref{fig:fric}d).

To calculate the excess work $w^{\rm FM}_{\rm ex}\equiv w^{\rm FM}-W_{\rm qs}$ for each trajectory, we subtract the discretized version of the quasistatic work~\eqref{ss-work} from the external work $w^{\rm FM}$~\eqref{st_work}. Averaging over $\mathcal{N}_{\rm R}=10^6$ independent realizations gives the average excess work $W^{\rm FM}_{\rm ex}$ (Figs.~\ref{fig:ratio_NS_LR} and \ref{fig:excess_works}). We compute the normalized flux (Fig.~\ref{fig:flux}) similarly.

\subsection{Simulation parameters}
\label{s_para}
For each numerical simulation, we choose discretization time $\Delta t/\traptime = 8 \cdot 10^{-4}$ ($8\cdot 10^{-2}$ times the smallest value of $t_{\rm a}$) and set inverse temperature $\beta = 1$ and diffusion constant $D=1$.

%$\bibliographystyle{unsrt}
%\bibliography{ref}


\end{document}